\documentclass[preprint,review,12pt]{elsarticle}

\usepackage{amssymb}
\usepackage{amsthm}
\usepackage{eqnarray}
\usepackage{lineno}
\usepackage{nomencl}
\usepackage{multicol}
\usepackage{framed}
\usepackage{ifthen}
\makenomenclature
\renewcommand*\nompreamble{\begin{multicols}{2}}
\renewcommand*\nompostamble{\end{multicols}}
\renewcommand{\nomgroup}[1]{%
  \vspace{0.6em}%
  \item[\itshape%
    \ifthenelse{\equal{#1}{A}}{Latin letters}{%
    \ifthenelse{\equal{#1}{G}}{Greek symbols}{%
    \ifthenelse{\equal{#1}{S}}{Subscripts}{%
    \ifthenelse{\equal{#1}{X}}{Superscripts}{}}}}%
  ]%
}

\newcommand\Pran{\mbox{\textit{Pr}}} 

\usepackage[colorlinks=true,bookmarks=false,citecolor=blue,urlcolor=blue]{hyperref} 
\setcitestyle{square}
\usepackage{amsmath}
\usepackage{tikz}
\usepackage{graphicx}
\graphicspath{Image}
\graphicspath{float}
\usepackage{caption}
\usepackage{subcaption}
\usepackage{setspace}
\usepackage{tabularx}
\usepackage{multirow}
\usepackage{rotating}
\usepackage{hyperref}
\usepackage{longtable}
\hypersetup{
     linkcolor = blue,
     urlcolor    = blue
}

\journal{Applied Thermal Engineering}

\begin{document}

\begin{frontmatter}



\title{Streamwise periodic solver to simulate turbulent flow in finned channels with uniform isothermal walls.}


\author[inst1,inst2]{Nitish Anand}
\ead{nitish.anand@vito.be}
\affiliation[inst1]{organization={Flemish Institute for Technological Research (VITO)},
            addressline={Boertang 200}, 
            city={Mol},
            postcode={3600}, 
            country={Belgium}}

\author[inst1,inst2]{Praharsh Pai Raikar}

\author[inst1,inst2,inst3]{Carlo De Servi}

\affiliation[inst2]{organization={EnergyVille},
            addressline={Thor Park 8310}, 
            city={Genk},
            postcode={3600}, 
            country={Belgium}}
\affiliation[inst3]{organization={Delft University of Technology},
            addressline={Kluyverweg 1}, 
            city={Delft},
            postcode={2629HS}, 
            country={The Netherlands}}


\begin{abstract}
Assessment of the thermo-hydraulic performance of heat exchangers using computational fluid dynamics is a challenging task. The intricate geometries of a heat exchanger require a fine discretization of the flow passage, which consequently leads to high computational costs. A streamwise periodic flow model can significantly reduce this cost, particularly for heat exchangers featuring repeating structures. This manuscript presents the streamwise-periodic turbulent source terms for flows in channels with isothermal walls, along with the implementation of the corresponding periodic flow solver in the open-source CFD-Suite, SU2. The accuracy of the implemented solver was verified by comparing its predictions against those of a full fin array simulation for the test case of offset circular fins. The results show that the streamwise periodic flow solver accurately reproduces the solutions of the full array simulation under both laminar and turbulent flow conditions.
\end{abstract}

\begin{graphicalabstract}
\centering
\includegraphics[width=0.45\linewidth, trim={6cm 21.8cm 6.45cm 1.5cm}, clip]{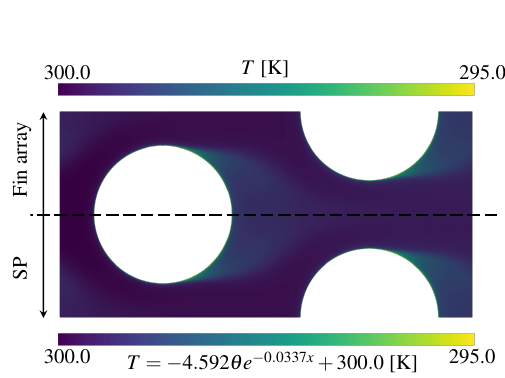}
\includegraphics[width=0.45\linewidth, trim={6cm 21.8cm 6.45cm 1.5cm}, clip]{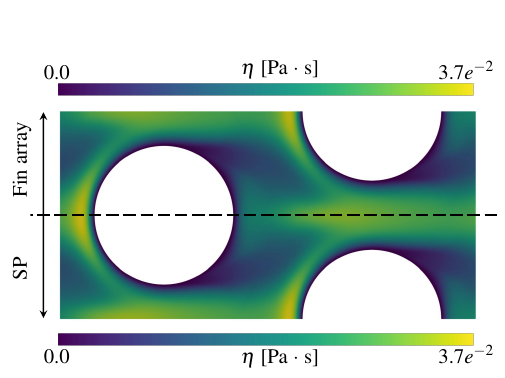}
\newline
\textsc{SP: streamwise periodic}
\end{graphicalabstract}

\begin{highlights}
\item Derivation of iso-thermal source terms for streamwise periodic flow solver
\item Verification of periodic quantities with full-array simulation
\item Entrance effects are more pronounced for turbulent flow
\end{highlights}


\end{frontmatter}


 \section{Introduction}

The thermal design of heat exchangers~(HEX) is generally carried out based on simplified methods such as $\epsilon$-NTU~(number of transfer units) and LMTD~(logarithmic mean temperature difference) methods \cite{BookShah}, complemented with empirical correlations for the prediction of heat transfer coefficients. Similarly, the hydraulic performance of the HEX is currently evaluated using correlations deduced from experimental results. These methods are well established for standard HEX types, for example, shell-and-tube, plate and fin-and-plate or fin-tube HEXs. To account for the variation of the fluid properties, the HEX is discretized in a few control volumes or cells. The heat transfer coefficient and pressure drop are calculated based on the local average fluid properties in each cell. Then, the performance of the entire HEX is estimated by combining the results obtained for all control volumes. Due to the uncertainty in the estimation of the heat transfer coefficient and friction factor, along with the simplifications introduced by the modeling approach, a conservative safety factor is applied in the HEX sizing process to ensure design feasibility. This inevitably leads to a sub-optimal design of the HEX.

To address this issue, HEX design using computational fluid dynamics~(CFD) has recently gained popularity within the scientific community. CFD simulations are now commonly used to characterize the thermo-hydraulic performance of HEXs~\cite{Muhammad2012}. Additionally, CFD models are combined with optimization methods for design purposes. For example, in Ref.~\cite{Josh2022}, the authors presented a shape optimization method to design the passages of a plate-type HEX. The proposed method relies on free-form deformation~(FFD) boxes to parametrize the channel geometry. Furthermore, the works in Ref.~\cite{Feppon2019, Ole2022} demonstrated the potential of shape optimization techniques in combination with computational models encompassing both the fluid and structure domains. The solution of a conjugate heat transfer problem also underpins topology optimization methods. Relevant studies focusing on the topological optimization of HEXs include, for example, Ref.~\cite{Pietropaoli2018, Lukas2020}.

Despite the effectiveness of these optimal design techniques, their application is largely confined to simplified test cases or to the design of a limited portion of the overall HEX geometry. This is primarily due to the high computational cost associated with the flow simulation in the case of complex geometries with multiple parallel channels. More specifically, to accurately predict the flow and heat transfer characteristics in channels with intricate shapes, a large number of mesh elements is required in the flow domain. The consequence is that the computational cost of the simulations is substantially high. This occurs although the CFD model, typically, is limited to only one side of a HEX — either the hot or cold side — and the heat transfer between the two streams is modeled by specifying Dirichlet or Neumann-type boundary conditions at the interface that separates them. The Dirichlet boundary condition, which specifies a constant temperature, is usually applied in the case of evaporators or condensers, as the temperature in the medium undergoing a phase change remains relatively constant, except in cases involving zeotropic mixtures. The Neumann boundary condition, which implies applying a constant heat flux, is instead considered when the temperature difference between the hot and cold streams tends to remain constant along the flow direction, as in the case of recuperators.

A common approach in the literature to reduce the computational burden associated with optimizing HEX geometry is surrogate model-based optimization. This methodology involves the development of a surrogate for the CFD model of the heat transfer device based on data obtained through simulations of the original high-fidelity model. The surrogate is then employed in the optimization process to estimate objective functions and constraints with significantly reduced computational time, although at the cost of increased simulation uncertainty. Notable examples of surrogate model-based optimization applied to HEX design are the studies in Ref.~\cite{Paola2014, Bacellar2017, Wen2016EnergyMOGAshellTube, Liu2017MOGAplate-fin, MANN2019MOGA2Dmicrofins}.

An effective alternative to reduce the computational cost associated with the simulation of HEXs, while also avoiding the time-consuming process of generating the dataset to calibrate the surrogate model, is the use of reduced order modeling techniques such as the streamwise periodic~(SP) method. This modeling technique has gained interest in the HEX community owing to its simplicity and the significant computational cost benefits. SP models are based on the assumption that in HEXs with repeating geometrical patterns, the flow properties become self-similar after a certain distance from the inlet. For instance, in the case of a finned-plate HEX, this characteristic of the flow and temperature fields can be exploited to reduce the size of the computational domain from the entire plate with its fin arrays to just a single fin and the flow domain around it. To solve the flow and heat transfer process in this elementary part or unit cell of the HEX, with dimensions corresponding to the pitch distances between the fins and the channel height, the Navier-Stokes equations are complemented with additional source terms and are efficiently solved using standard numerical methods.

The SP modeling technique was first proposed by \textit{Patankar et al.}~\cite{Patankar1977} for laminar flows. The authors derived the momentum and thermal source terms of the Navier-Stokes equations for both isothermal and heat-flux boundary conditions at the channel walls. Reference~\cite{GeertJFM_2016} extended the method to conjugate heat transfer in channels with periodically repeating fin structures, while Ref.~\cite{Wang2017} presents a SP flow solver for the laminar flow regime and isothermal wall boundaries based on the Lattice Boltzmann method. The application of SP solvers for thermo-hydraulic analysis of finned channel topologies in the case of laminar flow regime and isothermal walls was also demonstrated in Ref.~\cite{Steven2005} for channels with offset strip fins, and in Ref.\cite{Buckinx_Baelmans_2015} for pin-finned plates.

Although SP flow solvers are well-established for the laminar flow regime, their use in the case of turbulent flows remains limited. Reference~\cite{StalioDNS} reports the results of a thermo-hydraulic analysis performed with a SP direct numerical simulation~(DNS) solver to characterize a wavy channel with isothermal walls in laminar and transitional flow regimes. Besides, Ref.~\cite{tobias-phd} provides the source terms of the Navier-Stokes equations for SP turbulent flows in channels with prescribed heat-flux at the walls. However, to the authors' knowledge, no work in the open literature presented the mathematical formulation of the Navier-Stokes equations for simulating turbulent flows in channels with isothermal walls.

The present work documents the derivation of a SP Reynolds-Average Navier-Stokes (SP-RANS) solver for isothermal wall boundary conditions and its implementation in the open-source CFD-Suite \textit{SU2}~\cite{SU2}. To verify the SP solver, its predictions for a pin-fin unit cell were compared with those from the \textit{SU2} RANS solver for an equivalent full pin-fin array. The pin fins are arranged in a staggered layout, and the simulations cover both the laminar and turbulent flow regimes.

The paper is structured into four sections. Firstly, in the \textit{methodology} section, the source terms of the incompressible SP-RANS solver are derived for isothermal wall boundary conditions. Next, in the \textit{case study}, the numerical setup, fin geometry, boundary conditions and the thermophysical properties used for CFD simulations are detailed. In the \textit{results} section, results from the SP solver and standard RANS solver are compared. Finally, the \textit{conclusion} section summarizes the key takeaways from the research.
\section{Methodology}
After recalling the main governing equations, this section presents the derivation of the source term in the energy equation of the SP flow solver for the case of isothermal wall boundary conditions.

\subsection{Incompressible Navier-Stokes equations}
\label{sec:IncNSEqn}
The Navier-Stokes equations for viscous, incompressible flows at low-Mach number (pressure work negligible) under the assumption of constant fluid properties and low Eckert number (effect of viscous dissipation on fluid temperature negligible) can be written in their conservative form as 
\begin{eqnarray}
    \label{eq:NSEqn}
    \frac{\partial \mathcal{V}}{\partial t} + \nabla \cdot \mathcal{F}^c(\mathcal{V}) - \nabla \cdot \mathcal{F}^v(\mathcal{V}, \nabla \mathcal{V}) + \mathcal{S} = 0.
\end{eqnarray}
In Eq.~\eqref{eq:NSEqn} $\mathcal{V}$ is a set of conservative variables given as
\begin{eqnarray}
    \mathcal{V} = \left[ \begin{array}{c}
         \cdot  \\
         \rho \mathbf{u}  \\
         \rho c_{p} T
    \end{array}\right],
\end{eqnarray}
where $\rho$ is the fluid density, $\mathbf{u}$ is the velocity vector, $c_p$ is the specific heat at constant pressure and $T$ is the temperature. Further in Eq.~\eqref{eq:NSEqn}, $\mathcal{F}^c$ represents the convective fluxes and is given by
\begin{eqnarray}
    \mathcal{F}^c = \left[\begin{array}{c}
         \mathbf{u}  \\
         \rho \mathbf{u} \bigotimes \mathbf{u} + \bar{\bar{I}} p \\
         \rho c_p T \mathbf{u}
    \end{array}\right],
\end{eqnarray}
where $p$ is the pressure and $\bar{\bar{I}}$ is the identity matrix. Besides, $\mathcal{F}^v$ in Eq.~\eqref{eq:NSEqn} denotes viscous fluxes given as
\begin{eqnarray}
    \mathcal{F}^v = \left[\begin{array}{c}
         \cdot \\
         \bar{\bar{\tau}}\\
         \kappa \nabla T
    \end{array}\right].
\end{eqnarray}
 where $\kappa$ is the thermal conductivity of the fluid and $\bar{\bar{\tau}}$ is the viscous stress tensor. For an incompressible fluid, this term can be expressed as
 \begin{eqnarray}
 \label{eq:stressTensor}
     \bar{\bar{\tau}} = \mu \left(\nabla \mathbf{u} + \nabla \mathbf{u}^\top\right),
 \end{eqnarray}
 where $\mu$ is the dynamic viscosity.
 
 Finally, $\mathcal{S}$ in Eq.~\eqref{eq:NSEqn} is a generic source term which can be decomposed into its components as
 \begin{eqnarray}
     \mathcal{S} = \left[\begin{array}{c}
          \mathcal{S}_\mathrm{mass}\\
          \mathcal{S}_\mathrm{mom}\\
          \mathcal{S}_\mathrm{egy}\\
     \end{array}\right],
 \end{eqnarray}
representing the mass, momentum and energy source terms, respectively.

\subsection{Turbulence modeling}
 Turbulent flows are simulated by solving the Reynolds-Averaged Navier-Stokes (RANS) equations using the standard modeling approach based upon the Boussinesq hypothesis~\cite{pope_2000}. According to this hypothesis, the Reynolds stresses are proportional to the mean flow velocity gradients through the turbulent viscosity, denoted by $\mu_\textrm{turb}$. Thus, the effect of turbulence can be represented as an increase in viscosity, and the viscosity in the stress tensor, see Eq.~\eqref{eq:stressTensor}, can be expressed as
 \begin{eqnarray}
     \mu = \mu_\textrm{fluid} + \mu_\textrm{turb}.
 \end{eqnarray}

Similarly, the decomposition of velocity and temperature into mean and fluctuating components in the energy equation introduces a turbulent heat flux term, whose contribution is assumed proportional to the mean temperature gradient via the turbulent thermal diffusivity. As for the viscosity, the effective thermal conductivity for turbulent flows includes two terms
\begin{eqnarray}
    \kappa = \frac{\mu_\textrm{fluid} c_p}{\text{Pr}} + \frac{\mu_\textrm{turb} c_p}{\text{Pr}_\textrm{turb}},
\end{eqnarray}
where Pr represents the Prandtl Number.
The turbulent viscosity and Prandtl number terms are obtained from the turbulence models as specified in Ref.~\cite{Tom2018, Ole2019}.

For more details on the incompressible RANS equations and the solution method implemented in the open source CFD-Suite \textit{SU2}, interested readers are referred to Ref.~\cite{Tom2018, Ole2019}, while information on the available boundary conditions for SP solver can be found in Ref.~\cite{tobias-phd}.

\subsection{Streamwise periodic flow equations}
A fully developed flow under the assumption of constant fluid properties is considered SP, if the fluid dynamic variables, like velocity and pressure, after a certain distance from the inlet, do not depend on the streamwise coordinate~\cite{Patankar1977}. Hence, a flow field can be termed as SP over a distance L if it satisfies the following conditions
\begin{eqnarray}
    \label{eq:v_swp}
    \mathbf{u} (\mathbf{x})  &=& \mathbf{u} (\mathbf{x} + {\text{L}}), \\
    \label{eq:p_swp}
    p (\mathbf{x})  &=& p (\mathbf{x} + {\text{L}}) + \Delta p(\mathbf{x}),
\end{eqnarray}
where, $\mathbf{x}$ is the position in the flow domain, $p$ is the pressure, and $\Delta p$ is the pressure drop per unit cell. Substituting Eqs.~\eqref{eq:v_swp}~and~\eqref{eq:p_swp} into the incompressible Navier-Stokes flow equations, reported in~\autoref{sec:IncNSEqn}, leads to the source term for the momentum equation
\begin{eqnarray}
    \mathcal{S}_\mathrm{mom} = - \frac{\Delta p}{\text{L}}.
\end{eqnarray}
The derivation of this source term has been documented previously in Ref.~\cite{GeertJFM_2016, tobias-phd, Harikrishnan2020} and is reported in this work for completeness.

\subsection{Streamwise periodic temperature equation}
\label{EnergyEquation}
The formulation of the energy equation of a SP solver depends on the type of thermal boundary condition prescribed at the walls, namely, fixed heat-flux or isothermal condition. The source terms corresponding to the heat flux boundary condition have already been reported in Ref.~\cite{tobias-phd}. This section deals with the derivation of the source terms for the energy equation of a unit cell with isothermal walls and turbulent flow.

To facilitate the modeling of heat transfer in ducts with periodically varying cross-section and uniform wall temperature, Patankar \textit{et al.} introduced the concept of dimensionless reduced temperature $\theta$~\cite{Patankar1977}, defined as
\begin{eqnarray}
    \label{eqn:theta}
	\theta(\mathbf{x}) = \frac{T (\mathbf{x}) - T_\mathrm{w}}{T_\mathrm{b}({x}) - T_\mathrm{w}},
\end{eqnarray}
where $T$ is the local fluid temperature, $T_\mathrm{b}$ is the bulk temperature, which varies only in the streamwise flow direction ($x$) and is yet to be determined, and $T_\mathrm{w}$ is the wall temperature. From a physical point of view, $\theta$ represents the fluid temperature field relative to the wall (or fin) temperature, normalized by a reference or bulk temperature $T_\mathrm{b}$ such that $\theta$ varies periodically in the flow domain
\begin{eqnarray}
    \label{eqn:theta_periodic}
	\theta(\mathbf{x}) = \theta(\mathbf{x}+\mathrm{L}).
\end{eqnarray}

At the same time, the temperature difference between the wall and the flow decays exponentially to zero along the length (streamwise direction) of a duct with isothermal walls, see Ref.~\cite{Shah1974, StalioDNS}. If this temperature difference is expressed in terms of the flow bulk temperature, its decay rate can be defined as
\begin{eqnarray}
    \label{eqn:lambdaL}
    \lambda(\mathbf{x}) = -\frac{d (T_\mathrm{b}(\mathbf{x}) - T_\mathrm{w}) / d\mathbf{x}}{T_\mathrm{b}(x) - T_\mathrm{w}}.
\end{eqnarray} 

Equations \eqref{eqn:theta} and \eqref{eqn:lambdaL} together describe the temperature field in a thermally developed periodic domain. More specifically, Eq.~\eqref{eqn:theta} defines the normalized temperature distribution within a unit cell of a periodic domain, whereas Eq.~\eqref{eqn:lambdaL} defines the temperature difference decay along the streamwise direction.

Building on this, Patankar~\textit{et al.} \cite{Patankar1977} proved that, if the variation of the flow cross-section is periodic in the streamwise direction and the flow is fully developed, then $\lambda$ is periodic. It follows that the average value of $\lambda(\mathbf{x})$ is uniform and independent of the position $\mathbf{x}$. This can be expressed as
\begin{eqnarray}
    \label{eqn:lambda_L_a}
    \lambda_\text{L} = \frac{1}{\text{L}} \int^{\mathbf{x}+\text{L}}_{\mathbf{x}} \lambda (\mathbf{x}) d\mathbf{x}.
\end{eqnarray}

Next, substituting $\lambda (\mathbf{x})$ in  Eq.~\eqref{eqn:lambda_L_a} with the expression in Eq.~\eqref{eqn:lambdaL} and integrating over the periodic length L, it follows that
\begin{eqnarray}
    \label{eqn:temp_decay}
    (T_\mathrm{b} (\mathbf{x} + \text{L}) - T_\mathrm{w}) = (T_\mathrm{b} (\mathbf{x}) - T_\mathrm{w} )~e^{-\lambda_\text{L} \text{L}}.
\end{eqnarray}

According to Eq. \eqref{eqn:temp_decay}, the difference between the fluid bulk temperature and that of the wall decays exponentially over successive unit cells. The reduced temperature defined in Eq.~\eqref{eqn:theta} can therefore be written as 
\begin{eqnarray}
    \label{eqn:theta_bar}
    {\theta}(\mathbf{x}) = \frac{T(\mathbf{x})-T_\mathrm{w}}{(T_\mathrm{b} (x_0)-T_\mathrm{w})e^{-\lambda_\text{L} (x_0+ x)}},
\end{eqnarray}
and has the favorable property of being periodic in the streamwise direction~\cite{StalioDNS}. Note that $x_0$ indicates the streamwise location of the reference cell.

Expressing the energy equation in terms of this reference periodic temperature yields 
\begin{eqnarray}
    \label{eq:EnergyEqn}
	\frac{\partial}{\partial t} \theta + \nabla  \cdot (\theta \mathbf{u}) = \alpha \nabla^2 \theta + \mathcal{S}_\mathrm{egy},
\end{eqnarray}
where $\alpha$ is the thermal diffusivity ($\alpha=\frac{\kappa}{\rho c_p}$). Neglecting the possible time-dependency of $T_\mathrm{b}$, the formulation of the source term $\mathcal{S}_\mathrm{egy}$ varies depending on the flow regime as
\begin{eqnarray}
    \label{eq:EnergySourceLam}
    \mathcal{S}_\mathrm{egy,lam} &=& \alpha \lambda_\text{L}^2\theta  + \left(\mathbf{u} \cdot \hat{i_\mathrm{L}}\right)\lambda_\text{L} \theta  - 2 \alpha \lambda_\text{L} \left( \hat{i_\mathrm{L}} \cdot \nabla \theta \right), \\
    \label{eq:EnergySourceTurb}
    \mathcal{S}_\mathrm{egy,turb} &=& \mathcal{S}_\mathrm{egy,lam} - \theta \lambda_\text{L} \frac{\nabla \mu_\mathrm{turb} \cdot \hat{i_\mathrm{L}}}{\rho \text{Pr}_\mathrm{turb}},
\end{eqnarray}
where $\hat{i_\mathrm{L}}$ represents unit vector in the streamwise periodic direction.

The expression for the laminar source term $\mathcal{S}_\mathrm{egy,lam}$ has been well documented in Ref.~\cite{Buckinx_Baelmans_2015, StalioDNS, OF_SWP}; Conversely, to the authors' knowledge, the derivation of the turbulent source term $\mathcal{S}_\mathrm{egy,turb}$ constitutes a novel contribution of this work and has not yet been documented in the open literature. $\text{Pr}_\mathrm{turb}$ is assumed to be constant, as is common in most turbulence models. Moreover, for these flow conditions, the diffusivity term in equations \ref{eq:EnergySourceLam} and \ref{eq:EnergySourceTurb}, as in the energy equation, includes the turbulent thermal diffusivity contribution.

%
%
%
%

The solution of the energy equation ~\eqref{eq:EnergyEqn} requires the computation of $\lambda_\text{L}$. This quantity can be computed by integrating the energy equation over the unit cell~\cite{StalioDNS} under the steady-state assumption. Given the periodicity of $\theta$ and $\mathbf{u}$ and the fact that both vanish at the walls, the integration yields a quadratic equation of the form
\begin{eqnarray}
    A \lambda_\text{L}^2 + B \lambda_\text{L} + C = 0,
\end{eqnarray}
where
\begin{eqnarray}
    A &=&  \int_v \alpha \theta dv, \\
    B &=& \int_v \left( \left(\mathbf{u} \cdot \hat{i_\mathrm{L}}\right) \theta - \frac{\nabla \mu_\mathrm{turb} \cdot \hat{i_\mathrm{L}}}{\rho \Pran_\mathrm{turb}} \theta - 2 \alpha \left( \hat{i_\mathrm{L}} \cdot \nabla \theta \right) \right) dv,\\
    C &=& \int_s \alpha \left( \nabla \theta \cdot \Vec{n} \right) ds. 
\end{eqnarray}

\noindent It follows that the value of $\lambda_\text{L}$ can be determined from

\begin{eqnarray}
    \lambda_\text{L} = \frac{ - B + \sqrt{B^2 - 4AC}}{2A}.
\end{eqnarray}

\noindent After solving for the $\theta$ field, the solution is normalized by its volume average quantity to keep the solution stationary, as proposed in Ref.~\cite{Patankar1977, Wang2017, Harikrishnan2020}.

The source terms and the solution method for $\lambda_\text{L}$ were implemented in the open-source CFD-Suite \textit{SU2}~\cite{SU2,Tom2018}.

\section{Case Study}
\label{sec:CaseStudy}
\begin{figure}[t]
\centering
\includegraphics[width=0.5\linewidth, trim={6cm 23.5cm 6cm 3.5cm}, clip]{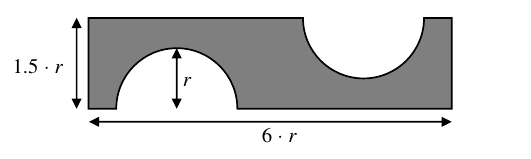}
\caption{Illustration of the unit cell flow domain representing the offset pin-fin array, where $r$ is the radius of the pin-fins.}
\label{fig:1cyl_illustration}
\end{figure}

To verify the implemented source terms, a representative test case is simulated: an offset pin-fin array, whose arrangement is defined by a longitudinal pitch of $6\cdot r$ and a lateral pitch of $1.5 \cdot r$, see~\autoref{fig:1cyl_illustration}, with $r$ indicating the radius of the pin-fins.

\begin{figure}[t]
\centering
\includegraphics[width=0.9\linewidth,trim={4.5cm 22cm 4.5cm 4.5cm}, clip]{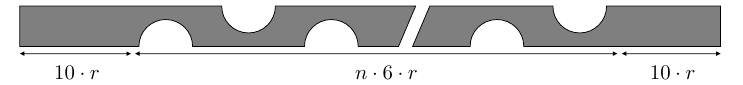}
\caption{Illustration of the simulation domain considered for the offset pin-fin array, where $r$ is the radius of the pin-fins.}
\label{fig:15cyl_illustration}
\end{figure}

The flow through the pin-fin array is simulated using two approaches: (i) the streamwise periodic flow solver, whose governing equations were derived in~\autoref{sec:IncNSEqn}, and (ii) a standard RANS solver applied to a domain defined by $n$ unit cells in series (see the flow domain illustrated in \autoref{fig:15cyl_illustration}). The corresponding boundary conditions are listed in \autoref{tab:bcs}. The value of the parameter $n$ is selected such that the flow in the final section of the fin array is fully periodic for the prescribed flow conditions.

To independently verify the laminar and turbulent source terms, two flow conditions were simulated, corresponding to Reynolds numbers of $100$~(laminar flow) and $10,000$~(turbulent flow). The adopted Reynolds number definition is 
\begin{eqnarray}
\text{Re} = \frac{\rho \cdot u_{\mathrm{L}} \cdot 2 \cdot r}{\mu}.
\end{eqnarray}
The flow properties corresponding to these flow conditions are reported in~\autoref{tab:cases}.

Regarding the adopted discretization methods, the convective fluxes were evaluated using a flux-difference-splitting scheme~\cite{Toro1997}, with second-order accuracy achieved via a MUSCL approach~\cite{Tom2018}, except for the convective fluxes corresponding to the turbulent quantities, which were computed using a first-order scalar upwind scheme. The viscous fluxes were evaluated using a corrected average-gradient method \cite{scheme1}, while the gradients and flow variables needed to evaluate the convective and viscous fluxes were calculated using the weighted least-squares approach \cite{scheme2}.

The flow domain was discretized using a structured grid of quadrilateral elements. The thickness of the first element of the inflation layer around the fins was selected to ensure a $y^+$ close to unity, enabling accurate capture of the flow characteristics close to the wall. The thermophysical properties were modeled based on the ideal gas assumptions, while the turbulence quantities were evaluated using the k-$\omega$ shear stress transport turbulence model~\cite{sst2003}. The flow solution was obtained by resorting to the Euler implicit time-marching scheme with CFL numbers of $40$ and $100$ for the streamwise periodic and RANS simulations, respectively. A six order of magnitude reduction in the residuals of the mass, momentum, and energy equations was used as the convergence criterion.

\begin{table}[t]
\centering
\caption{Boundary conditions used for the simulation of the laminar and turbulent flow test cases.}
\begin{tabular}{ccccccc}
\hline
\multirow{2}{*}{Cases}   & \multicolumn{1}{c}{$T_\mathrm{in}$}    & \multicolumn{1}{c}{$\dot{m}_\mathrm{in}$}  & \multicolumn{1}{c}{$\mathcal{I}$} & \multicolumn{1}{c}{$\sigma$} & \multicolumn{1}{c}{$p_\mathrm{out}$}& \multicolumn{1}{c}{$T_\mathrm{w}$}       \\
   & \multicolumn{1}{c}{[K]}    & \multicolumn{1}{c}{[kgm$^{-3}$]}  & \multicolumn{1}{c}{$[-]$} & \multicolumn{1}{c}{$[-]$} & \multicolumn{1}{c}{[Pa]}& \multicolumn{1}{c}{[K]}   \\
\hline
\hline
Laminar  & 288.15  & 0.75 & - & - & 0.0 & 300.0\\
Turbulent & 288.15 & 0.75 & 0.1 & 15.0 & 0.0 & 300.0\\
\hline
\end{tabular}
\label{tab:bcs}
\end{table}

\begin{table}[t!]
\centering
\caption{Flow properties corresponding to  the laminar and turbulent test cases.}
\begin{tabular}{ccccccc}
\hline
\multirow{2}{*}{Cases}   & \multicolumn{1}{c}{$Re$}   & \multicolumn{1}{c}{$u_x$}     & \multicolumn{1}{c}{$\rho$}  & \multicolumn{1}{c}{$r$} & \multicolumn{1}{c}{$\mu$}  & \multicolumn{1}{c}{Pr}       \\
   & \multicolumn{1}{c}{[$-$]}   & \multicolumn{1}{c}{[{ms$^{-1}$}]}     & \multicolumn{1}{c}{[{kgm}$^{-3}$]}  & \multicolumn{1}{c}{[m]} & \multicolumn{1}{c}{[kgm$^{-1}${s}$^{-1}$]} & \multicolumn{1}{c}{[$-$]}      \\
\hline
\hline
Laminar  & 100  & 1.0 & 1.0 & 0.5 & 0.01 & 0.72 \\
Turbulent & 10000 & 1.0 & 1.0 & 0.5 & 0.0001 & 0.90\\
\hline
\end{tabular}
\label{tab:cases}
\end{table}

\subsection{Grid Independence Study}
\begin{table}[ht]
\centering
\caption{Metrics of the grid convergence study for the four meshes considered in the laminar and turbulent flow test cases, where $\epsilon$ represents the relative uncertainty in the mesh.}
\begin{tabular}{cc|llll|llll}
\hline
\multirow{2}{*}{S. No.} & \multirow{2}{*}{$\xi_i$} & \multicolumn{4}{c|}{Laminar} & \multicolumn{4}{c}{Turbulent}                                                         \\ 
    \ & & \multicolumn{1}{c}{$\lambda_\mathrm{L}$} & \multicolumn{1}{c}{$\epsilon$\%} & \multicolumn{1}{c}{$\Delta p$}     & \multicolumn{1}{c|}{$\epsilon$\%} & \multicolumn{1}{c}{$\lambda_\mathrm{L}$} & \multicolumn{1}{c}{$\epsilon$\%} & \multicolumn{1}{c}{$\Delta p$}     & \multicolumn{1}{c}{$\epsilon$\%} \\ \hline
\hline
1 & 1997.3   & \multicolumn{1}{c}{2.321e-1} & \multicolumn{1}{c}{0.94} & \multicolumn{1}{c}{6.47} & \multicolumn{1}{c|}{0.95}  & \multicolumn{1}{c}{2.51e-2} & \multicolumn{1}{c}{17.1} & \multicolumn{1}{c}{4.40} & \multicolumn{1}{c}{30.56}     \\ 
2 & 8359.0  & \multicolumn{1}{c}{2.343e-1} & \multicolumn{1}{c}{0.21} & \multicolumn{1}{c}{6.42} & \multicolumn{1}{c|}{0.19}  & \multicolumn{1}{c}{3.04e-2} & \multicolumn{1}{c}{9.8} & \multicolumn{1}{c}{3.37} & \multicolumn{1}{c}{4.01}     \\ 
3 & 26092.8  & \multicolumn{1}{c}{2.358e-1} & \multicolumn{1}{c}{0.04} & \multicolumn{1}{c}{6.41} & \multicolumn{1}{c|}{0.03}  & \multicolumn{1}{c}{3.37e-2} & \multicolumn{1}{c}{0.30} & \multicolumn{1}{c}{3.24} & \multicolumn{1}{c}{0.09}     \\ 
4 & 43314.7  & \multicolumn{1}{c}{2.359e-1} & \multicolumn{1}{c}{-} & \multicolumn{1}{c}{6.40} & \multicolumn{1}{c|}{-}  & \multicolumn{1}{c}{3.38e-2} & \multicolumn{1}{c}{-} & \multicolumn{1}{c}{3.24} & \multicolumn{1}{c}{-}     \\ \hline
\end{tabular}
\label{tab:GCI}
\end{table}

To estimate the discretization error, a mesh convergence study was performed in accordance with the guidelines in Ref.~\cite{ASME-GCI}. For both the SP-solver  laminar and turbulent flow test cases, two performance metrics were considered, namely the temperature decay exponent ($\lambda_\mathrm{L}$) and the pressure drop ($\Delta p$). Four meshes featuring 2926, 12246, 38226, and 63456 elements, hereafter referred to as coarse, medium, fine, and extra-fine, respectively, were examined. All meshes were generated using the open-source software \textit{gmsh}~\cite{gmsh,pygmsh}.

\begin{figure}[ht!]
\centering
    \begin{subfigure}{0.49\textwidth}
        \includegraphics[width=1\textwidth]{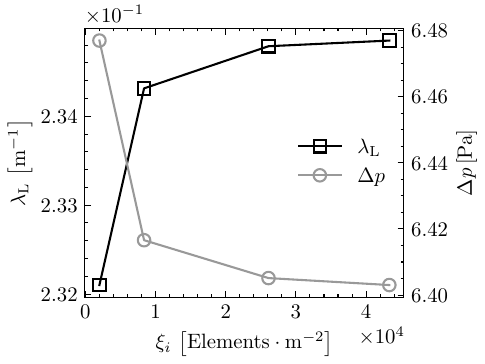}
        \caption{}
    \end{subfigure}
    \begin{subfigure}{0.49\textwidth}
            \includegraphics[width=1\textwidth]{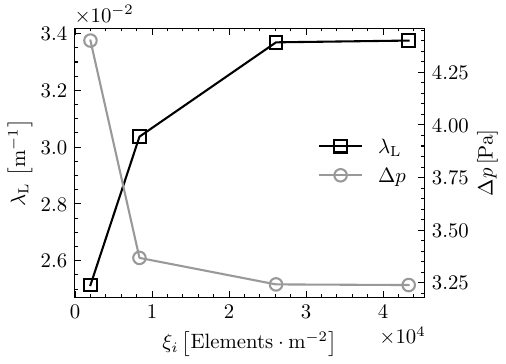}
        \caption{}
    \end{subfigure}
    \caption{Variation of the metrics of interest ($\lambda_\mathrm{L}$ \& $\Delta p$) with grid density ($\xi_{i}$), in the case of (a)~laminar (Re$ = 100$) and (b)~turbulent (Re$ = 10000$) flows.}
    \label{fig:GCI}
\end{figure}

\begin{figure}[ht!]
    \centering
    \includegraphics[width=0.5\linewidth, trim={25cm 15cm 25cm 15cm}, clip]{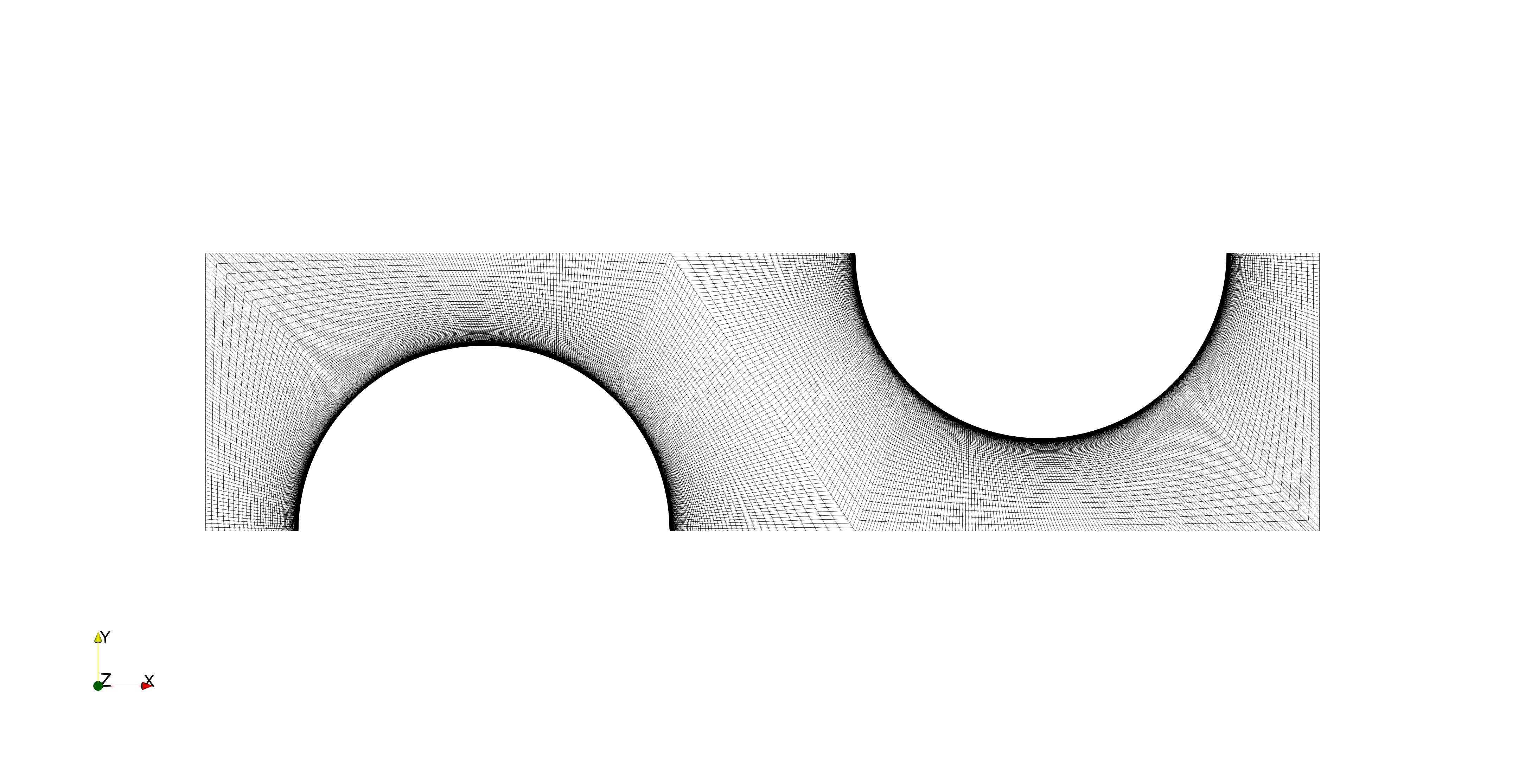}
    \caption{Mesh selected for the numerical simulations.}
    \label{fig:mesh}
\end{figure}

\autoref{fig:GCI} illustrates the variation of the two metrics of interest,  $\lambda_\text{L}$ and ${\Delta p}$, with grid density ($\xi_i$), defined as the number of elements per unit area. Monotonic convergence of both metrics was observed for the two flow regimes. \autoref{tab:GCI} reports the values of $\lambda_\text{L}$ and ${\Delta p}$ obtained using the four meshes. These results indicate that the predictions obtained with the fine mesh deviate by less than 0.3\% from the values corresponding to the extra-fine mesh. Hence, the fine mesh resolution is deemed sufficient for the numerical study reported in this work. The topology of the fine mesh is illustrated in~\autoref{fig:mesh}. To ensure consistent accuracy in the SP and RANS simulations, the mesh density determined from the mesh convergence study for the SP solver was retained for the fin array simulations.

\section{Results}
In the following sections, the results obtained using the streamwise periodic flow solver are compared with those from simulations of an entire fin array performed with standard \textit{SU2} solvers, in order to verify the implementation of the laminar and turbulent source terms derived in \autoref{sec:IncNSEqn}. First, the laminar test case is examined, followed by a comparison of the turbulent streamwise periodic flow solver (SP-RANS) results with those of the RANS solver.

\subsection{Laminar Flow Regime ($\text{Re} = 100$)}
The simulations performed with the SP solver encompassed only a fin element, as shown in~\autoref{fig:1cyl_illustration}, with uniform temperature assumed at the fin surface (isothermal boundary condition). Conversely, the computational domain in the simulations with the standard \textit{SU2} solver included 10 fin elements arranged in series ($n$=11), see \autoref{fig:15cyl_illustration}. The boundary conditions and the flow properties prescribed in these simulations are reported in \autoref{tab:bcs} and \autoref{tab:cases}, respectively.

\subsubsection{Identifying flow streamwise periodicity along the fin array}
\begin{figure}[ht]
\centering
\includegraphics[width=\linewidth]{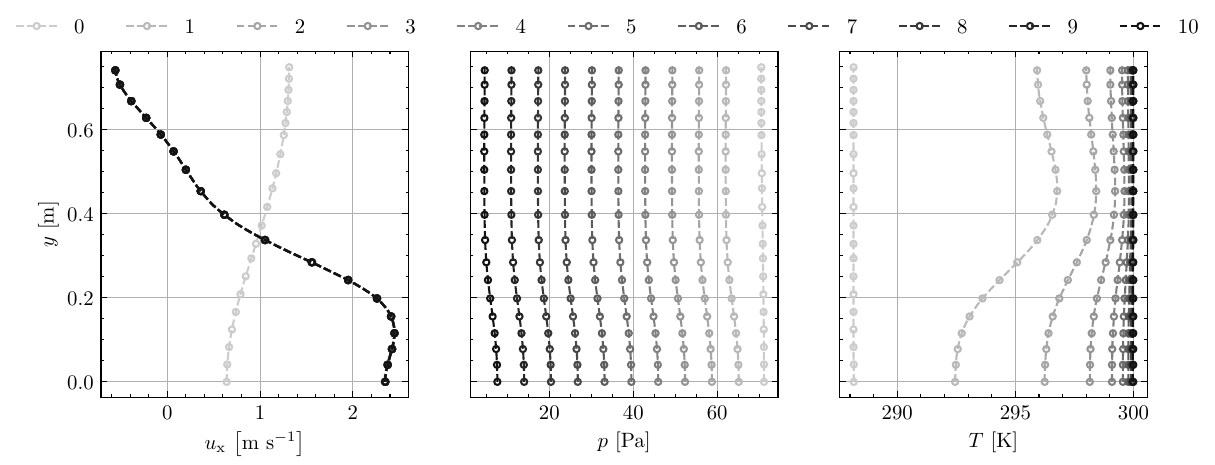}
\caption{Variation of flow properties at the inlet of each unit cell according to the results of the fin array simulation for $Re$=100: streamwise velocity (left), pressure (center) and temperature (right).}
\label{fig:NumericalVerification_100_a}
\end{figure}

Before comparing the flow and temperature fields predicted using the two solvers, it is necessary to identify the fin element from which the flow becomes streamwise periodic. To this end, \autoref{fig:NumericalVerification_100_a} reports the flow streamwise velocity, pressure, and temperature at the inlet of each unit cell according to the results of the fin array simulation. It can be observed that the streamwise velocity profile becomes identical beyond unit cell \#2, see the left chart of ~\autoref{fig:NumericalVerification_100_a}. In addition, the pressure variation per unit cell is constant beyond unit cell \#3, as shown in the chart in the middle of ~\autoref{fig:NumericalVerification_100_a}. Furthermore, as expected, the temperature difference between the fluid and the walls decreases exponentially from unit cell \#2 to \#8, see the right panel of \autoref{fig:NumericalVerification_100_a}.

\begin{figure}[ht]
\centering
\includegraphics[width=0.5\linewidth]{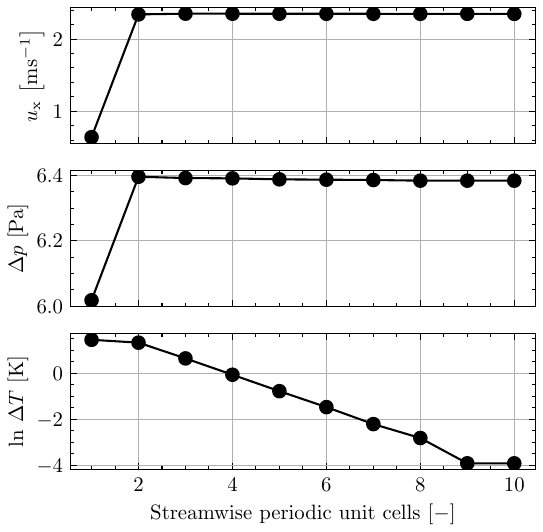}
\caption{Variation of the streamwise flow velocity (top chart), pressure drop across each unit cell (middle), and logarithm of the fluid temperature difference between the inlet and outlet of each unit cell (bottom), as obtained from the fin array simulation for $Re$= 100.}
\label{fig:NumericalVerification_100_b}
\end{figure}

These observations are confirmed by the trends in~\autoref{fig:NumericalVerification_100_b}, whose upper and middle charts display the values of streamwise velocity and pressure at the center of the inlet plane of each unit element forming the fin array. 
The change in streamwise velocity is negligible beyond unit cell \#2, while $\Delta p$ becomes constant beyond unit cell \#4. Besides, the bottom graph of \autoref{fig:NumericalVerification_100_b} illustrates the evolution of $\ln(\Delta T)$ along the fin array, where $\Delta T$ is the temperature difference between the inlet and outlet of each unit cell. As expected, the temperature difference decays exponentially, consistently with the theory of thermally developed flows~\cite{Patankar1977}, from unit cell \#4 to \#9. 

Therefore, it can be concluded that the flow is fluid-dynamically and thermally developed between unit cells \#5-\#8. These conditions are not fully established in unit cell \#9 due to the strong influence of exit effects in the last unit cell of the fin array on the upstream flow field. Based on the results presented, the flow and temperature fields estimated for unit cell \#5 were selected for the comparison with the solution obtained using the streamwise periodic flow solver.

%

\subsubsection{Verification of the SP solver for laminar flows}
\begin{figure}[ht]
     \centering
     \begin{subfigure}[b]{0.49\textwidth}
         \centering
         \includegraphics[width=\textwidth, trim={6cm 21.8cm 6.45cm 1.5cm}, clip]{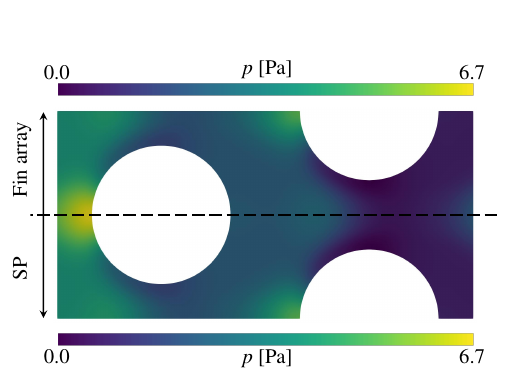}
         \caption{Pressure}
     \end{subfigure}
     \begin{subfigure}[b]{0.49\textwidth}
         \centering
         \includegraphics[width=\textwidth, trim={6cm 21.8cm 6.45cm 1.5cm}, clip]{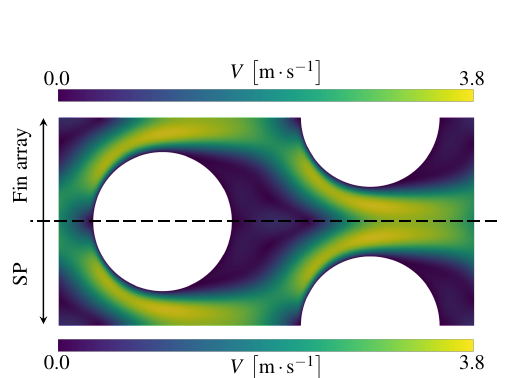}
         \caption{Velocity Magnitude}
     \end{subfigure}
     \begin{subfigure}[b]{0.49\textwidth}
         \centering
         \includegraphics[width=\textwidth, trim={6cm 21.8cm 6.45cm 1.5cm}, clip]{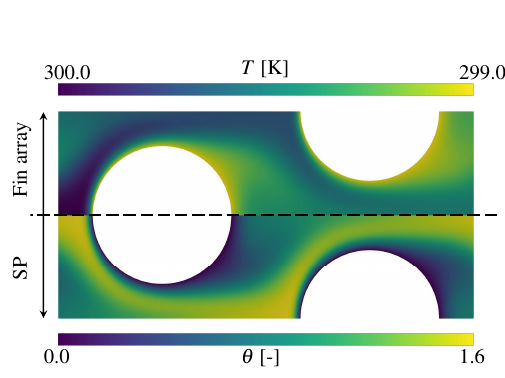}
         \caption{Temperature}
     \end{subfigure}
     \begin{subfigure}[b]{0.49\textwidth}
         \centering
         \includegraphics[width=\textwidth, trim={6cm 21.8cm 6.45cm 1.5cm}, clip]{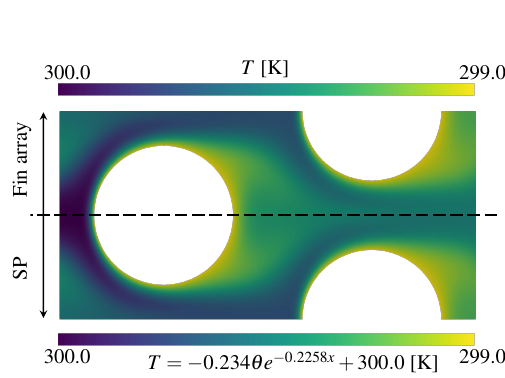}
         \caption{Reconstructed Temperature}
     \end{subfigure}
        \caption{Comparison of the results obtained from the fin array simulation with those from the SP solver.}
        \label{fig:flow_field_100}
\end{figure}

In this section, the flow and temperature fields of unit cell \#5, as estimated from the fin array simulation, are compared with those obtained from the streamwise periodic flow solver. This comparison is shown in \autoref{fig:flow_field_100}. It can be observed that the pressure and velocity fields determined with the two solvers are identical. For the temperature distribution, the solution is first compared with the flow reduced temperature ($\theta$) determined by the SP-NS solver, see ~\autoref{fig:flow_field_100}(c). Note that the $\theta$ distribution is periodic in the streamwise direction, as defined in Eqn.~\eqref{eqn:theta}, differently from the actual flow temperature. The $\theta$ field can be used to reconstruct the actual temperature distribution given the value of $\lambda_\text{L}$, as per Eqn.~\eqref{eq:theta_scale}. The comparison between the reconstructed temperature field and that obtained from the array simulation is illustrated in \autoref{fig:flow_field_100}(d). It is apparent that the two fields correspond perfectly.

\begin{figure}[ht!]
	\centering
	\includegraphics[width=0.50\linewidth, trim={6cm 22cm 6cm 2cm}, clip]{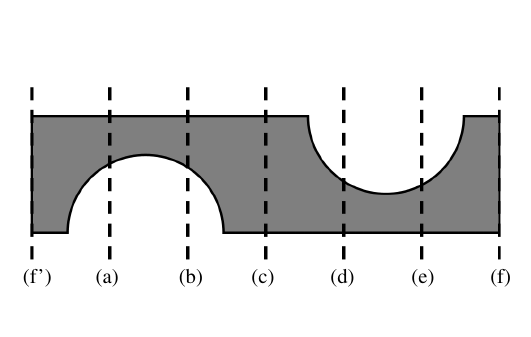}
    \caption{Illustration of the sections where flow properties are sampled for the comparison of the solutions from the standard flow solver and from the streamwise periodic flow solver.} 
    \label{fig:planes}
\end{figure}
\begin{figure}[ht]
\centering
\includegraphics[width=\linewidth]{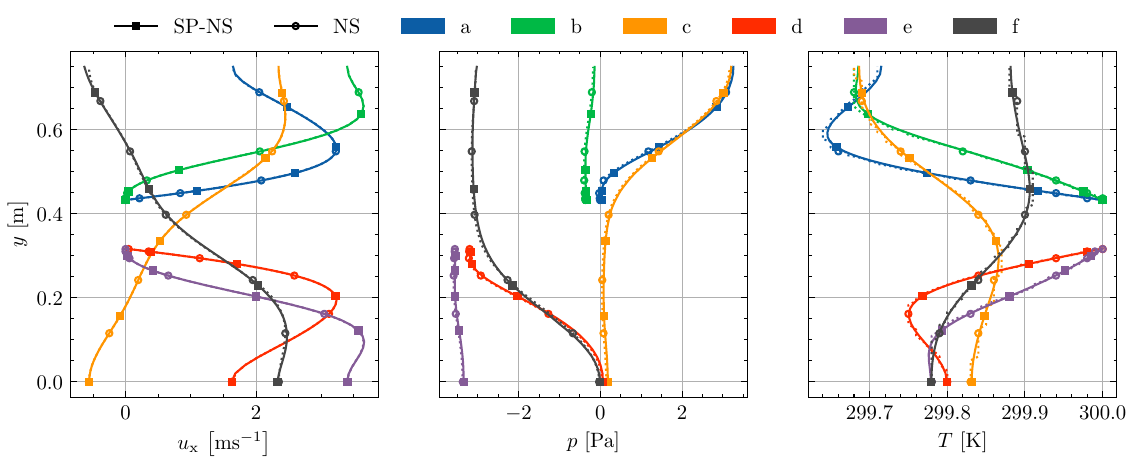}
\caption{Comparison of the solution from the streamwise periodic flow solver with that from the fin array simulation for laminar flow conditions ($Re=100$). Labels (a)-(f) indicate the positions in the flow domain, as shown in \autoref{fig:planes}, at which the flow properties have been sampled. Solid black lines represent the results from the streamwise periodic solver, while dotted grey lines represent the results obtained from the simulation of the fin array with a conventional solver.}
\label{fig:NumericalVerification_100}
\end{figure}

To enable a quantitative comparison, flow properties from both solutions are sampled along six equidistant lines (a-f) perpendicular to the streamwise direction. The  positions of these lines are indicated in~\autoref{fig:planes}, and the respective flow properties are plotted in \autoref{fig:NumericalVerification_100}. Note that lines (f') and (f) correspond to the inlet and outlet sections of the unit cell, respectively. Moreover, the scaled flow properties calculated by the SP solver are converted to actual flow quantities. Specifically, the pressure values from the SP solver are offset by the back-pressure of unit cell \#5 in the fin array simulation. The reduced temperature from the SP solver ($\theta$) is converted to absolute temperature using
\begin{eqnarray}
\label{eq:theta_scale}
    T = D_\text{1}\theta \exp^{-\lambda_\text{L} x} +~D_\text{2}.
\end{eqnarray}
Here, $\lambda_\text{L}$ is the temperature difference exponential decay rate, $x$ indicates the location in the streamwise direction in the unit cell domain, and $D$ are constants. The values of $\lambda_\text{L}$ and $\Delta p$ computed by the SP solver are 0.2258 $\mathrm{m}^{-1}$ and 6.41 Pa, respectively, as also tabulated in~\autoref{tab:SP-values}. 
It is apparent from \autoref{fig:NumericalVerification_100} that the temperature, velocity, and pressure profiles from the simulation with the SP solver align well with those obtained for unit cell \#5 in the fin array simulation with the standard incompressible solver of \textit{SU2}. Hence, it can be concluded that the expression derived in this work for the source term of the energy equation of the SP solver, along with its implementation, are accurate for simulating laminar flows.

\begin{table}[t!]
\centering
\caption{Main flow characteristics as predicted by the SP flow solver for the laminar and the turbulent flow conditions considered in the study.}
\begin{tabular}{ccccc}
\hline
Cases   & {$\dot{m}$}   & {$\lambda_\mathrm{L}$}     & {$\Delta p$} & {$Eu$} \\
   & {[$\mathrm{kg} \mathrm{m}^{-2} \mathrm{s}^{-1}$]}   & {[$\mathrm{m}^{-1}$]}     & {[Pa]} & {[-]} \\
\hline
\hline
Laminar  & 0.75  & 0.2258 & 6.41 & 6.41 \\
Turbulent & 0.75 & 0.0337 & 3.24 & 3.24 \\
\hline
\end{tabular}
\label{tab:SP-values}
\end{table}
\subsection{Turbulent Flow Regime ($Re = 10000$)}
This subsection reports the results of the verification study assessing the accuracy of the implemented energy equation source term in the SP solver, see Eq.~\eqref{eq:EnergySourceTurb}, for turbulent flows. The verification procedure is analogous to that followed in the previous section for laminar flow conditions: the results obtained using the SP flow solver are compared against those from a simulation of an entire fin array performed with a standard RANS solver. The computational domains of the two simulations are identical to those of the laminar test case, while the corresponding flow properties and boundary conditions are reported in \autoref{tab:cases} and \autoref{tab:bcs}, respectively.

\subsubsection{Identifying flow streamwise periodicity along the fin array}
\begin{figure}[ht]
\centering
\includegraphics[width=\linewidth]{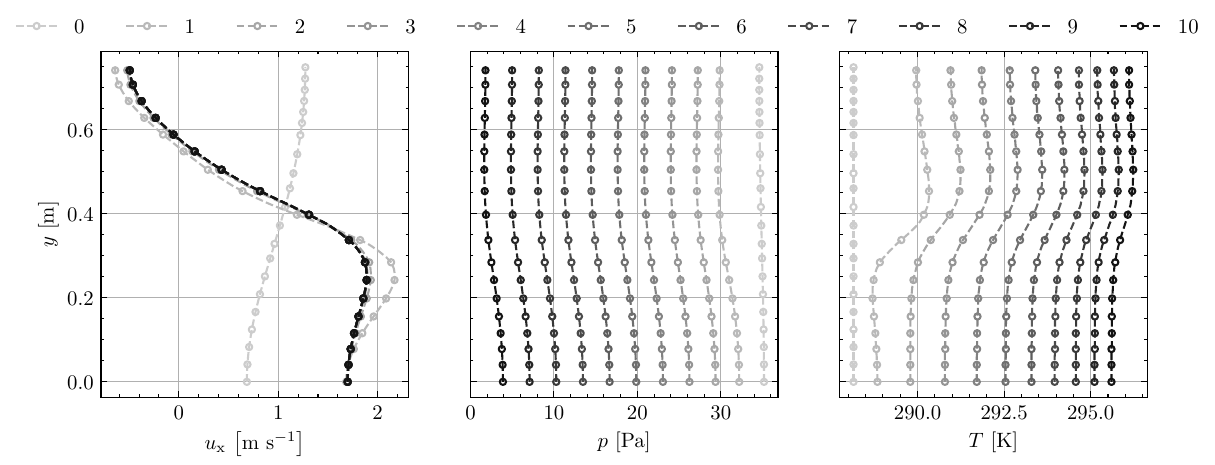}
\caption{Variation of flow properties at the inlet of each unit cell according to the results of the RANS simulation for $Re$=10000: streamwise velocity (left), pressure (center), and temperature (right).}
\label{fig:NumericalVerification_5000_a}
\end{figure}

\autoref{fig:NumericalVerification_5000_a} reports the flow streamwise velocity, pressure, and temperature at the inlet of each unit element based on the results of the RANS solver. Compared to the laminar test case, the streamwise velocity distribution becomes identical further downstream of the inlet, namely beyond unit cell \#4, as shown in the left chart of \autoref{fig:NumericalVerification_5000_a}. Similarly, the change in pressure after each unit cell is constant beyond unit cell \#6, as seen in the center graph of \autoref{fig:NumericalVerification_5000_a}. In contrast, the temperature profile tends to be self-similar already after unit cell \#2, as in the laminar test case, see the right chart of \autoref{fig:NumericalVerification_5000_a}.
%

\begin{figure}[ht]
\centering
\includegraphics[width=0.5\linewidth]{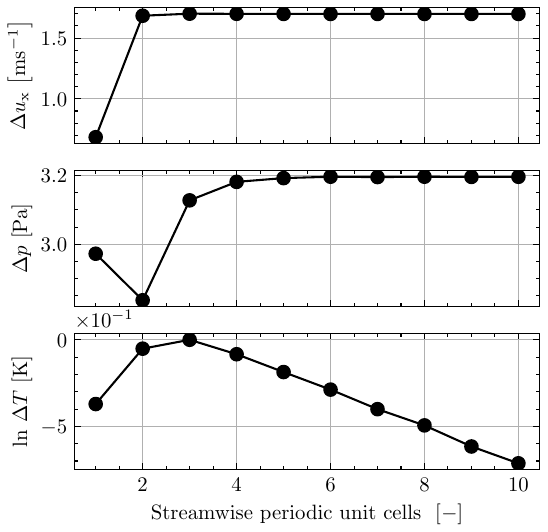}
\caption{Variation of the streamwise flow velocity (top chart), pressure drop across each unit cell (middle), and logarithm of the fluid temperature difference between the inlet and outlet of each unit cell based on the results of the RANS solver for $Re$=10000.}
\label{fig:NumericalVerification_5000_b}
\end{figure}

To better identify where streamwise periodicity is attained, flow quantities sampled at the center of the plane corresponding to the inlet of each unit cell in the fin array simulation are displayed in \autoref{fig:NumericalVerification_5000_b}. As shown in the top graph of \autoref{fig:NumericalVerification_5000_b}, the velocity is essentially constant beyond unit cell \#4. The trend in the chart at the center of \autoref{fig:NumericalVerification_5000_b} confirms the observation made for \autoref{fig:NumericalVerification_5000_a}, specifically that the pressure drop ($\Delta p$) becomes constant beyond unit cell \#6. This indicates that the flow is fully developed in the region downstream of unit cells \#5. Moreover, the right chart of \autoref{fig:NumericalVerification_5000_b} shows that the logarithm of the temperature difference between the inlet and outlet of each unit cell ($\ln\Delta T$) decays linearly between unit cells \#4-\#8. Hence, it can be concluded that the flow is thermally developed beyond cell \#4. It follows that the portion of the RANS solution most suitable for a comparison with the results of the SP solver is that spanning unit cells \#7-\#9. In particular, the flow field of unit cell \#8 was selected for the comparison.

\subsubsection{Verification of the SP solver for turbulent flows}
\begin{figure}[ht!]
     \centering
     \begin{subfigure}[b]{0.49\textwidth}
         \centering
         \includegraphics[width=\textwidth, trim={6cm 21.8cm 6.45cm 1.5cm}, clip]{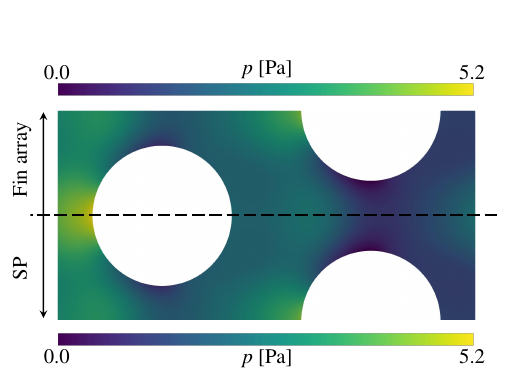}
         \caption{Pressure}
     \end{subfigure}
     \begin{subfigure}[b]{0.49\textwidth}
         \centering
         \includegraphics[width=\textwidth, trim={6cm 21.8cm 6.45cm 1.5cm}, clip]{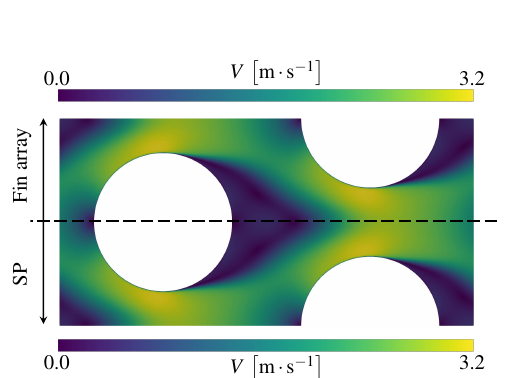}
         \caption{Velocity Magnitude}
     \end{subfigure}
     \begin{subfigure}[b]{0.49\textwidth}
         \centering
         \includegraphics[width=\textwidth, trim={6cm 21.8cm 6.45cm 1.5cm}, clip]{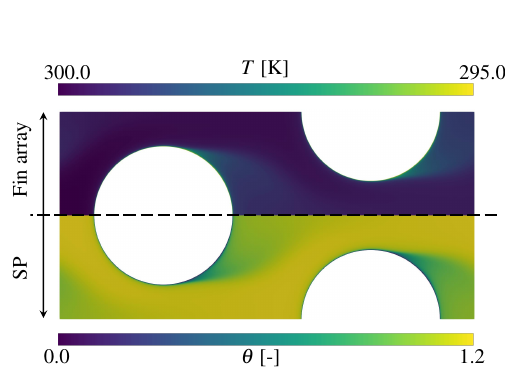}
         \caption{Temperature}
     \end{subfigure}
      \begin{subfigure}[b]{0.49\textwidth}
         \centering
         \includegraphics[width=\textwidth, trim={6cm 21.8cm 6.45cm 1.5cm}, clip]{Image/RE5000/SIM/Temperature_reconstructed_5000.pdf}
         \caption{Reconstructed Temperature}
     \end{subfigure}
     \begin{subfigure}[b]{0.49\textwidth}
         \centering
         \includegraphics[width=\textwidth, trim={6cm 21.8cm 6.45cm 1.5cm}, clip]{Image/RE5000/SIM/EddyViscosity_5000.pdf}
         \caption{Eddy Viscosity}
     \end{subfigure}
        \caption{Comparison of the RANS solver solution with that from the SP solver.}
        \label{fig:flow_field_5000}
\end{figure}

The solution from the RANS simulation for unit cell \#8 is compared with the results obtained using the SP solver in~\autoref{fig:flow_field_5000}. As shown in ~\autoref{fig:flow_field_5000}(a) and (b), the pressure and streamwise velocity distributions estimated through the two solvers are essentially identical. \autoref{fig:flow_field_5000}(c) displays the temperature field from the fin array simulation alongside the reduced temperature ($\theta$) from the SP solver, which is periodic in the streamwise direction, as defined in Eq.~\eqref{eqn:theta_bar}. As for the laminar test case, the temperature field can be reconstructed from the $\theta$ distribution given the value of $\lambda_\mathrm{L}$, as defined in Eq.~\eqref{eq:theta_scale}. The comparison between the temperature field from the RANS solver simulation and the reconstructed temperature distribution based on the SP solver solution is shown in \autoref{fig:flow_field_5000}(d). It is apparent that the two temperature fields are identical. The same consideration applies to the eddy viscosity ($\eta$) fields from the two solvers, which are compared in \autoref{fig:flow_field_5000}(e).

\begin{figure}[ht!]
\centering
\includegraphics[width=\linewidth]{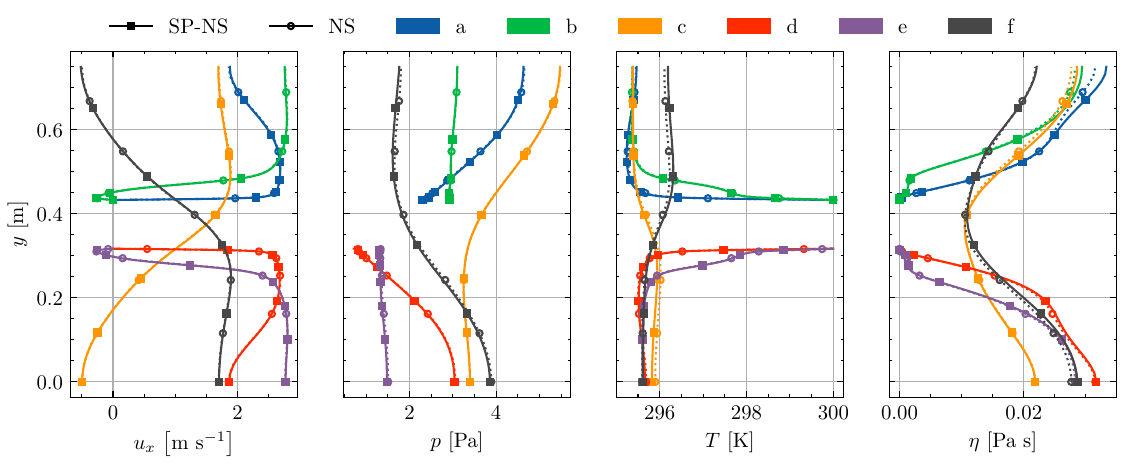}
\caption{Comparison of the solution from the SP solver with that from the fin array simulation for turbulent flow conditions ($Re=10000$). Labels (a)-(f) indicate the section in the flow domain, as represented in \autoref{fig:planes}, at which the flow properties
have been sampled. The solid black lines represent the results from the SP solver, while the dotted grey lines represent the results obtained from the simulation of the flow through the fin array with the RANS solver.}
\label{fig:NumericalVerification_5000}
\end{figure}

To enable a quantitative comparison, flow properties from both solutions are sampled along six equidistant cross-sections, positioned as illustrated in \autoref{fig:planes}. The comparison between the solutions of the two solvers is shown in \autoref{fig:NumericalVerification_5000}. The scaled temperature and pressure computed by the SP solver have been converted to actual flow quantities using the same procedure applied in the laminar test case.
The value of $\lambda_\text{L}$ needed for this purpose is reported in \autoref{tab:SP-values}. It can be observed that the trends in velocity, pressure, and temperature obtained from the SP solver align closely with those from the RANS simulation. Minor deviations are seen only in the eddy viscosity. Hence, it can be concluded that the contribution due to turbulence in the source term derived in this work for the energy equation of the SP solver is accurate.

\subsection{Comparison of Laminar and Turbulent Flow Test Cases}
\autoref{tab:SP-values} reports the pressure drop ($\Delta p$) and the decay rate of the temperature difference between the flow and the walls ($\lambda_\text{L}$) across a unit cell, as estimated in the SP solver simulations. It is observed that $\lambda_\text{L}$ differs by nearly an order of magnitude between the laminar and turbulent flow conditions examined. Specifically, $\lambda_\text{L}$ is 0.23~$\mathrm{m}^{-1}$ for the laminar flow case and 0.03~$\mathrm{m}^{-1}$ for the turbulent case, indicating more effective heat transfer in the laminar flow regime. In contrast, $\Delta p$ in the laminar test case is approximately twice that of the turbulent simulation, with values of 6.41~Pa and 3.24~Pa, respectively. 

The computational cost of the fin array simulations performed in this study averaged $\sim$1440 minutes (about one day), while the corresponding unit fin simulations using the SP solver required $\sim$30 minutes on a workstation equipped with an \textit{Intel Xeon Gold 5220R} (2.2~GHz) processor and 192~GB of memory. This demonstrates the substantial computational savings that can be achieved by using an SP solver to simulate flow and heat transfer in finned plates or channels. Such a reduction in computational cost makes the optimization of complex heat exchanger topologies practically viable when using SP solvers.
\section{Conclusions}
The research reported in this manuscript presented the derivation of the source term of the energy equation for streamwise periodic flow simulation corresponding to iso-thermal boundaries, applicable to both laminar and turbulent flow regimes. The reported source term was implemented in the open-source CFD code \textit{SU2}. To verify the accuracy of the derived source term and its implementation, solutions from the streamwise periodic solver were compared with those obtained using a standard finite-volume solver, under both laminar ($Re=$ 100) and turbulent ($Re=$ 10000) flow conditions. The computational domain analyzed with the standard solver consisted of a channel containing an array of 11 staggered pin fins, whereas the streamwise periodic flow simulations employed a reduced domain corresponding to a single unit cell of the fin array. Based on the simulation results, the following conclusions can be drawn:
\begin{enumerate}
    \item The source term derived for the energy equation under the streamwise periodic flow assumption is shown to be accurate based on the results of a conventional solver for both laminar and turbulent flow conditions.
    \item Hydraulic and thermally developed flow conditions are attained at different locations along channels with repeating solid structures depending on the flow regime. For the case of offset circular fins, self-similar flow patterns were observed after fin \#5 under laminar conditions, while beyond fin \#7 for turbulent flow.
    \item The computational cost of streamwise periodic flow simulations for the test cases analyzed is $\sim$30 minutes, compared to $\sim$1440 minutes for the full fin array simulations. This substantial reduction in computational time demonstrates the advantage of employing streamwise periodic solvers in place of standard finite-volume solvers in optimization studies.
\end{enumerate}

Future work will focus on incorporating the developed streamwise periodic solver into an optimization framework.
    \nomenclature[A]{$\mathcal{F}$}{fluxes} 
    \nomenclature[A]{$\mathcal{V}$}{conservative variables} 
    \nomenclature[A]{$\mathcal{S}$}{source terms} 
    \nomenclature[A]{$A$}{coefficient} 
    \nomenclature[A]{$B$}{coefficient} 
    \nomenclature[A]{$C$}{coefficient} 
    \nomenclature[A]{$c$}{heat capacity} 
	\nomenclature[A]{$r$}{fin radius} 
    \nomenclature[A]{$D$}{constants} 
    \nomenclature[A]{$f$}{specific heat capacity}
    \nomenclature[A]{$\mathcal{I}$}{turbulence intensity} 
    \nomenclature[A]{$i$}{vector} 
    \nomenclature[A]{$I$}{identity matrix} 
    \nomenclature[A]{$\mathrm{L}$}{streamwise periodic pitch} 
    \nomenclature[A]{$m$}{mass} 
    \nomenclature[A]{$p$}{pressure} 
    \nomenclature[A]{$t$}{time} 
    \nomenclature[A]{$T$}{temperature} 
    \nomenclature[A]{$\mathbf{u}$}{velocity vector} 
    \nomenclature[A]{$v$}{volume} 
    \nomenclature[A]{u}{velocity component} 
    \nomenclature[A]{$x$}{x-axis} 
    \nomenclature[A]{$y$}{y-axis} 
    \nomenclature[A]{$y^{+}$}{first element thickness} 
    \nomenclature[A]{$n$}{number of fins} 
    \nomenclature[A]{$Pr$}{Prandtl Number} 
    \nomenclature[A]{$Re$}{Reynolds Number} 
    \nomenclature[A]{$Eu$}{Euler Number} 
    \nomenclature[G]{$\xi$}{mesh density} 
	\nomenclature[G]{$\eta$}{eddy viscosity} 
	\nomenclature[G]{$\Delta$}{difference in quantities} 
    \nomenclature[G]{$\lambda$}{temperature decay exponent} 
    \nomenclature[G]{$\theta$}{reduced temperature} 
    \nomenclature[G]{$\nabla$}{differential operator} 
    \nomenclature[G]{$\rho$}{density} 
    \nomenclature[G]{$\kappa$}{thermal conductivity} 
    \nomenclature[G]{$\alpha$}{thermal diffusivity} 
    \nomenclature[G]{$\mu$}{dynamic viscosity} 
    \nomenclature[G]{$\tau$}{viscous stress tensor} 
    \nomenclature[G]{$\sigma$}{eddy viscosity ratio} 
    \nomenclature[G]{$\epsilon$}{error} 
    \nomenclature[S]{egy}{energy} 
	\nomenclature[S]{mom}{momentum} 
    \nomenclature[S]{mass}{mass} 
    \nomenclature[S]{fluid}{fluid} 
    \nomenclature[S]{lam}{lam} 
    \nomenclature[S]{turb}{turbulent} 
    \nomenclature[S]{in}{inlet} 
    \nomenclature[S]{out}{outlet} 
    \nomenclature[S]{L}{periodic length} 
    \nomenclature[S]{w}{wall} 
    \nomenclature[S]{p}{pressure}
    \nomenclature[S]{i}{index} 
    \nomenclature[S]{b}{bulk} 
    \nomenclature[S]{$0$}{reference location} 
	\nomenclature[X]{$c$}{convective} 
	\nomenclature[X]{$v$}{viscous} 
    \nomenclature[X]{$\top$}{transpose} 
    \nomenclature[X]{$.$}{change in time} 
    \nomenclature[X]{$\hat{}$}{unit vector} 
     \printnomenclature
\section*{Declaration of competing interest}
The authors declare that they have no known competing financial interests or personal relationships that could have appeared to influence the work reported in this paper. 
\section*{Acknowledgement}
Research reported in this manuscript was funded by the Vlaams Agentschap Innoveren and Ondernemen (VLAIO), Belgium, through the project IAMHEX (grant number HBC.2021.0801). 



\appendix


 \bibliographystyle{elsarticle-num} 
 \bibliography{Bib/references}





\end{document}